\pdfoutput=1
\documentclass[twocolumn,superscriptaddress,aps,preprintnumbers,amsmath,amssymb,prd,nofootinbib]{revtex4}

\usepackage{amsmath}
\usepackage{amssymb}
\usepackage{amsfonts}
\usepackage{graphicx}
\usepackage{xcolor}
\usepackage{xfrac}
\usepackage{comment}
\usepackage{pifont}
\usepackage{physics}
\usepackage{fourier}
\usepackage{hyperref}
\usepackage{bm}
\usepackage{enumitem}
\usepackage{braket}

\definecolor{rossoferrari}{HTML}{D9073D}
\definecolor{mediumblue}{HTML}{0000CD}
\definecolor{forestgreen}{HTML}{228B22}
\definecolor{desy_blue}{HTML}{009EE2}
\definecolor{desy_orange}{HTML}{FD8800}
\definecolor{light_pink}{rgb}{1,0.4,0.4}
\definecolor{light_blue}{rgb}{0.284602,0.317763,0.963947}
\hypersetup{setpagesize=false,bookmarksnumbered=true,bookmarksopen=true,%
colorlinks=true,linkcolor=light_blue,urlcolor=rossoferrari,citecolor=rossoferrari,linktocpage=false}

\makeatletter
\renewcommand{\appendix}{\par
  \setcounter{section}{0}%
  \setcounter{subsection}{0}%
  \gdef\presectionname{\appendixname\,}%
  \gdef\postsectionname{}%
  \gdef\thesection{\presectionname\@Alph\c@section\postsectionname}%
  \gdef\thesubsection{\@Alph\c@section.\@arabic\c@subsection}%
  \renewcommand{\theequation}{\@Alph\c@section\arabic{equation}}%
  \renewcommand{\thefigure}{\@Alph\c@section.\arabic{figure}}%
  \renewcommand{\thetable}{\@Alph\c@section.\arabic{table}}%
}
\makeatother

\begin{document}

\title{Light Dark Matter Search with Nitrogen-Vacancy Centers in Diamonds}

\author{So Chigusa}
\affiliation{Berkeley Center for Theoretical Physics, Department of Physics,
University of California, Berkeley, CA 94720, USA}
\affiliation{Theoretical Physics Group, Lawrence Berkeley National Laboratory,
Berkeley, CA 94720, USA}

\author{Masashi Hazumi}
\affiliation{International Center for Quantum-field Measurement Systems for Studies of the Universe and Particles (QUP), High Energy Accelerator Research Organization (KEK), 1-1 Oho, Tsukuba, Ibaraki 305-0801, Japan}
\affiliation{Institute of Particle and Nuclear Studies (IPNS), KEK, Tsukuba, Ibaraki 305-0801, Japan}
\affiliation{Japan Aerospace Exploration Agency (JAXA), Institute of Space and Astronautical Science (ISAS), Sagamihara, Kanagawa 252-5210, Japan}
\affiliation{Kavli Institute for the Physics and Mathematics of the Universe (Kavli IPMU, WPI), UTIAS, The University of Tokyo, Kashiwa, Chiba 277-8583, Japan}
\affiliation{The Graduate University for Advanced Studies (SOKENDAI), Miura District, Kanagawa 240-0115, Hayama, Japan}

\author{Ernst David Herbschleb}
\affiliation{Institute for Chemical Research, Kyoto University, Gokasho, Uji-city, Kyoto 611-0011, Japan}

\author{Norikazu Mizuochi}
\affiliation{Institute for Chemical Research, Kyoto University, Gokasho, Uji-city, Kyoto 611-0011, Japan}
\affiliation{Center for Spintronics Research Network, Kyoto University, Uji, Kyoto 611-0011, Japan}
\affiliation{International Center for Quantum-field Measurement Systems for Studies of the Universe and Particles (QUP), High Energy Accelerator Research Organization (KEK), 1-1 Oho, Tsukuba, Ibaraki 305-0801, Japan}

\author{Kazunori Nakayama}
\affiliation{Department of Physics, Tohoku University, Sendai, Miyagi 980-8578, Japan}
\affiliation{International Center for Quantum-field Measurement Systems for Studies of the Universe and Particles (QUP), High Energy Accelerator Research Organization (KEK), 1-1 Oho, Tsukuba, Ibaraki 305-0801, Japan}

\begin{abstract}

We propose an approach to directly search for light dark matter, such as the axion or the dark photon, by using magnetometry with nitrogen-vacancy centers in diamonds. If the dark matter couples to the electron spin, it affects the evolution of the Bloch vectors consisting of the spin triplet states, which may be detected through several magnetometry techniques. We give several concrete examples with the use of dc and ac magnetometry and estimate the sensitivity on dark matter couplings. 
 
\end{abstract}


\maketitle

\section{Introduction}

The existence of dark matter in the Universe is a long-standing mystery of particle physics, astrophysics and cosmology.
Many experiments try to reveal the nature of dark matter, but it is not achieved yet~\cite{ParticleDataGroup:2022pth}.
One proposed candidate for dark matter is the axion, originally introduced to solve the strong CP problem in quantum chromodynamics~\cite{Peccei:1977hh,Weinberg:1977ma,Wilczek:1977pj}. 
Nowadays, a wider class of light bosonic dark matter models are frequently discussed, including axion-like particles and the dark photon. 
They lead to rich phenomenology and cosmology, and various search strategies have been proposed~\cite{Kawasaki:2013ae,Marsh:2015xka,Terrano:2019clh,DiLuzio:2020wdo,Jaeckel:2010ni,Graham:2015ouw,Caputo:2021eaa,Antypas:2022asj}, including an interesting approach using the $\mathrm{K}$--$^3\mathrm{H_e}$ comagnetometer~\cite{PhysRevX.13.011050}, as summarized in Ref.~\cite{OHare:2020}.

In this paper, we propose a new approach for detecting light bosonic dark matter by applying magnetometry with nitrogen-vacancy (NV) centers in diamonds~\cite{Degen:2017,Barry:2019sdg}. NV centers have drawn significant attention for their applications in diverse fields, from industry to bioscience, due to their precise magnetic sensing capabilities~\cite{Taylor:2008,Acosta:2009tr,Wolf:2015,Barry:2016,Barry:2019sdg,Du:2021}.
We exploit this property of NV centers to detect light bosonic dark matter, which couples to the electron spin and behaves as an effective magnetic field.\footnote{
  Applications of NV centers to particle physics have been considered in several contexts. Refs.~\cite{Rong:2017wzk,Chen:2019uai,Chu:2021agg} considered spin-dependent new forces acting on electrons. Their sensitive mass range and interaction strength for a new particle are both different from ours and also they are not related to dark matter. Ref.~\cite{Rajendran:2017ynw} considered directional detection of WIMP (weakly interacting massive particle) dark matter. The WIMP is much heavier than the dark matter candidates considered in this letter, and the detection method is completely different.
}
For example, axion-like dark matter $a$ has an interaction with the electron spin $\vec S_e$ through the effective Hamiltonian 
\begin{align}
	 H_{\rm eff} = \frac{g_{aee}}{m_e}\vec\nabla a\cdot\vec S_e,
\end{align}
and hence it is clear that the gradient of the axion looks like an effective magnetic field.
Our approach provides excellent sensitivity for these dark matter models, surpassing current observational bounds in certain cases.
We adopt natural units throughout the paper, $\hbar = c = 1$.

This paper is organized as follows.
In Sec.~\ref{sec:magnetometry} we review the basics of magnetometry with NV centers in diamonds. 
In Sec.~\ref{sec:DM} we give an idea for application of the magnetometry technique to the dark matter search. 
In Sec.~\ref{sec:dc} we estimate the sensitivity for axion or dark photon dark matter by using the DC magnetometry technique. 
In Sec.~\ref{sec:res} we estimate the sensitivity when the resonance happens.
In Sec.~\ref{sec:ac} we estimate the sensitivity by using the AC magnetometry technique. 
Sec.~\ref{sec:con} is devoted for conclusion and discussion.

\section{Magnetometry with NV center ensembles}
\label{sec:magnetometry}

We focus on the negatively charged state of the NV center, $\mathrm{NV}^-$, which is particularly suited for quantum sensing applications.
In this state, two electrons form a spin triplet, with three states represented by $\left|0\right>$ and $\left|\pm\right>$, corresponding to the spin along the $z$-axis $m_s=0$ and $\pm 1$, respectively. 
Denoting the sum of two spin operators as $\vec S_e = \vec S_{e1}+ \vec S_{e2}$, the Hamiltonian is given by~\cite{Barry:2019sdg}
\begin{align}
	H = 2\pi D S_{e,z}^2 + \gamma_e B_0 S_{e,z} + \gamma_e \vec B\cdot \vec S_e,
	\label{Hamiltonian}
\end{align}
where $D$ is the zero-field-splitting parameter, $B_0$ is an artificially-applied bias magnetic field along the $z$-axis, $\vec B = (B_x, B_y, B_z)^t$ is an external magnetic field to be sensed, and $\gamma_e\simeq e/m_e$ with $e$ and $m_e$ being the U(1) gauge coupling constant and the electron mass, respectively.
The energy difference between the ground state $\left|0\right>$ and $\left|\pm\right>$ is given by $\omega_\pm = 2\pi D \pm \gamma_e B_0$, where $D=2.87\,{\rm GHz}$ and $2\pi D = 11.9\,{\rm \mu eV}$.
With the magnetometry protocol described below, we can select and focus on two of these states.
We will consider the two-state system with $\left|0\right>$ and $\left|+\right>$, and study the time evolution of the Bloch vector, which is a superposition of these two states.

\subsection{DC magnetometry}

The dc magnetometry goes as follows~\cite{Barry:2019sdg}. 
(i) We prepare the initial state to be $\left|\psi\right>=\left|0\right>$ and apply a so-called $\pi/2$ pulse with the frequency equal to $\omega_+$.
(ii) This is followed by a free precession phase of duration $\tau$.
Usually for ensembles, $\tau$ is taken to be comparable to the spin dephasing time $T_2^* \sim 1\,\mathrm{\mu s}$~\cite{Barry:2016}.\footnote{
Precisely speaking, the dephasing time $T_2^*$ depends on the concentration of NV-centers.
In this paper, we neglect this dependence of the spin relaxation time for simplicity, as the number of NV centers can be changed both by concentration and by volume.
}
(iii) Finally, another $\pi/2$ pulse is applied, with the magnetic field direction tilted by an angle $\theta$ from the initial $\pi/2$ pulse. 
(iv) The final state is measured by the fluorescence light. 
This entire sequence is known as the Ramsey sequence~\cite{Ramsey:1950}.

Let us see time evolution of the Bloch vector in detail.
The interaction Hamiltonian is written as follows in the basis of triplet states $\left|+1\right>, \left|0\right>, \left|-1\right>$,
\begin{align}
	\gamma_e \vec B\cdot\vec S_e = \gamma_e \begin{pmatrix}
		B_z & \frac{1}{\sqrt 2} B_- & 0 \\
		\frac{1}{\sqrt 2} B_+ & 0 &  \frac{1}{\sqrt 2} B_- \\
		0 & \frac{1}{\sqrt 2} B_+ & -B_z \\
	\end{pmatrix},
\end{align}
where $B_\pm \equiv B_x \pm i B_y$. 
In the Ramsey protocol described below, a pulse with a frequency tuned to $\omega_+$ (or $\omega_-$) is applied and the $\left|+\right>$ (or $\left|-\right>$) state is selectively excited as described in detail below. Thus it is allowed to pick up and focus on two states, $\left|+\right>$ and $\left|0\right>$.
The Hamiltonian of this two level system, under the time varying magnetic field $\vec B(t)$, is written as
\begin{align}
	&H = H_0 + H_1,\\
	&H_0 = \frac{\omega_+}{2}\sigma_z,~~~~~~
	H_1 =  \gamma_e \begin{pmatrix}
		B_z & \frac{1}{\sqrt 2} B_-  \\
		\frac{1}{\sqrt 2} B_+ &  0
	\end{pmatrix},
\end{align}
after redefinition of the zero of the energy, where $\omega_+ = 2\pi D + \gamma_e B_0$.
The vector spanned by the basis $\left|+\right>$ and $\left|0\right>$ is called the Bloch vector.

To solve the Schr\"{o}dinger equation, it is convenient to go to the interaction picture where the state is redefined as $\left|\psi_I\right> = e^{iH_0 t} \left|\psi_S\right>$, where the subscript $I$ and $S$ denote the interaction and Schr\"{o}dinger picture, respectively. 
Time evolution of the state is governed by\footnote{
The effect of dephasing is not included in this evolution equation. It can be easily recovered later as the exponential factor in Eq.~\eqref{eq:proj_sensitivity}.
}
\begin{align}
	i \frac{\partial}{\partial t} \left|\psi_I(t)\right> = H_{I} \left|\psi_I(t)\right>,~~~~~~
	H_{I} = e^{i H_0 t} H_1 e^{-i H_0 t}.
\end{align}
Let us explain the Ramsey sequence consisting of four steps (i)-(iv) and how the Bloch vector evolves under this Hamiltonian~\cite{Degen:2017,Barry:2019sdg}.

\textit{(i) Initial pulse :} 
First, we prepare the initial state to be $\left|\psi_I(0)\right> = \left|0\right>$.
A pulse with $\vec B(t) = \vec B_1\cos(\omega t)$ with the polarization angle aligned to the $y$ direction is applied. 
Then the interaction Hamiltonian is written as
\begin{align}
	H_{I} = \frac{\gamma_e B_1}{2\sqrt{2}}
	\begin{pmatrix}
		0 & -i(e^{i(\omega_+ +\omega)t} + e^{i(\omega_+-\omega)t}) \\
		i(e^{-i(\omega_+ +\omega)t} + e^{-i(\omega_+-\omega)t}) & 0
	\end{pmatrix}.
\end{align}
For a pulse with $\omega\simeq \omega_+$, we can neglect the rapidly oscillating part $e^{\pm i (\omega_+ + \omega) t}$.
For the same reason, the off-diagonal elements of the interaction Hamiltonian between the $\ket{0}$ and $\ket{-}$ states, which we already dropped in the previous equation, can be neglected.
Thus, we can focus on the qubit system composed of $\ket{0}$ and $\ket{+}$ and obtain the effective time-evolution operator as
\begin{align}
	e^{-i H_{I} t} &\simeq \cos\left(\frac{\gamma_e B_1 t}{2\sqrt{2}}\right){\bf I} 
	-i\sigma_y\sin\left(\frac{\gamma_e B_1 t}{2\sqrt{2}}\right),
\end{align}
or equivalently, the time evolution of a state as
\begin{align}
	 \left|\psi_I(t)\right> 
	 = \left[\cos\left(\frac{\gamma_e B_1 t}{2\sqrt{2}}\right) {\bf I} -i  \sin\left(\frac{\gamma_e B_1 t}{2\sqrt{2}}\right) \sigma_y \right] \left|\psi_I(0)\right>.
\end{align}
For the so-called $\pi/2$ pulse, we choose $t_{\pi/2} = \pi/(\sqrt{2}\gamma_e B_1)$. 
For the initial condition $\left|\psi(0)\right> = \left|0\right>$, we have
\begin{align}
	 \left|\psi_I\left(t_{\pi/2}\right)\right> = \frac{1}{\sqrt 2}\left(-\left|+\right> + \left|0\right> \right).
\end{align}
Thus the state is maximally mixed after this initial pulse.

\textit{(ii) Free precession phase :} 
Next we consider an effect of external magnetic field $B(t)$, which we want to detect, on the state after the initial pulse is shut off. 
Taking $\tau$ as a duration of this phase, we obtain 
\begin{align}
	e^{-i \int_0^\tau H_I dt} = 
	\begin{pmatrix}
		e^{-i \phi(\tau)/2} & 0 \\
		0 & e^{i \phi(\tau)/2}
	\end{pmatrix},
\end{align}
where
\begin{align}
	\phi(\tau)  =  \int_0^\tau \gamma_e B_z(t) dt.
\end{align}
For nearly constant magnetic field, we obtain $\phi(\tau) \simeq \gamma_e B_z \tau$. 
As a result, the effect of the magnetic field on the state is to add a phase like
\begin{align}
	 \left|\psi_I\left(t_{\pi/2} +\tau \right)\right> 
	 &= \begin{pmatrix}
		e^{-i\phi(\tau)/2} &0 \\
		0 & e^{i\phi(\tau)/2}
	\end{pmatrix}
	 \left|\psi_I\left(t_{\pi/2}\right)\right>\\
	 &= \frac{1}{\sqrt 2}\left(-e^{-i\phi(\tau)/2}\left|+\right> + e^{i\phi(\tau)/2} \left|0\right> \right),
\end{align}
This extra phase $\phi$ acquired during the free precession phase contains information about the external magnetic field.
For our cases of interest, the magnetic field $B$ is so small that we can approximate $\phi \ll 1$.
In the next section we will consider an effective magnetic field induced by axion or dark photon dark matter.

\textit{(iii) Final pulse :} 
Finally we again apply a pulse with $\vec B=\vec B_2\cos(\omega t)$ and $\omega \simeq \omega_+$. The direction of the polarization of the final pulse measured from the $y$ direction is denoted by $\theta$. Then the interaction Hamiltonian, neglecting rapidly oscillating terms, is given by
\begin{align}
	H_{I} \simeq \frac{\gamma_e B_2}{2\sqrt{2}}\left(-\sigma_x \sin\theta  + \sigma_y\cos\theta \right).
\end{align}
In this case we obtain
\begin{align}
	e^{-i H_{I} t} &\simeq \cos\left(\frac{\gamma_e B_2 t}{2\sqrt{2}}\right){\bf I} 
	+i (\sigma_x\sin\theta-\sigma_y\cos\theta)\sin\left(\frac{\gamma_e B_2 t}{2\sqrt{2}}\right)\\
	&= \cos\left(\frac{\gamma_e B_2 t}{2\sqrt{2}}\right){\bf I}  + \begin{pmatrix}
		0 & -e^{-i\theta} \\
		e^{i\theta} & 0
	\end{pmatrix} \sin\left(\frac{\gamma_e B_2 t}{2\sqrt{2}}\right)\\
	&=\frac{1}{\sqrt 2}\begin{pmatrix}
		1 & -e^{-i\theta} \\
		e^{i\theta} & 1
		\end{pmatrix},
\end{align}
where we assumed $t=t_{\pi/2}$ in the last line.\footnote{
	A useful formula is $e^{i \vec X\cdot\vec\sigma} = {\bf I} \cos(X) + i (\hat X\cdot\vec\sigma) \sin(X)$ where $\vec X$ is a real vector and $X=|\vec X|$.}
After this final pulse, the final state is given by
\begin{align}
	\left|\psi_I(t)\right> &= U_2^{\pi/2}\, U_{\rm free}(\tau)\, U_1^{\pi/2} \begin{pmatrix}
		0 \\
		1
	\end{pmatrix} \notag \\
	&=-e^{-i\theta/2} \cos\left(\frac{\phi-\theta}{2}\right) \left|+\right>  + ie^{i\theta/2} \sin\left(\frac{\phi-\theta}{2}\right) \left|0\right>,
	\label{psi_fin_DC}
\end{align}
where $t = t_{\pi/2} + \tau + t_{\pi/2} \simeq \tau$ and
\begin{align}
	&U_1^{\pi/2} = \frac{1}{\sqrt 2}\begin{pmatrix}
		1 &  -1  \\
		1  & 1
	\end{pmatrix},~~~~~
	U_2^{\pi/2} = \frac{1}{\sqrt 2}\begin{pmatrix}
		1 &  -e^{-i\theta}  \\
		 e^{i\theta}  & 1
	\end{pmatrix},\\
	&U_{\rm free}(\tau) =\begin{pmatrix}
		e^{-i\phi/2} & 0  \\
		 0  & e^{i\phi/2}
	\end{pmatrix}.
\end{align}

\textit{(iv) Measurement of the state :}
Finally we measure the state after the final pulse, to obtain information about the magnetic field through the phase $\phi$. 
It is extracted by measuring the relative population between the states $\left|+\right>$ and $\left|0\right>$. To quantify it, we define
\begin{align}
	S \equiv \frac{1}{2} \left<\psi(\tau)\right| \sigma_z \left|\psi(\tau)\right>
        =\frac{1}{2}\cos(\phi-\theta).
        \label{S_fin_DC}
\end{align}
In particular, for $\theta= \pi/2$, we have
\begin{align}
	S = \frac{1}{2}\sin\phi.   \label{S_fin_DC_pi2}
\end{align}
For small $\phi$, $S$ is proportional to $B_z$. Thus the deviation of $S$ from zero signals the external magnetic field. 
Physically the measurement is done by looking at the fluorescence light, which reflects the relative population between $\left|+\right>$ and $\left|0\right>$.

\subsection{AC magnetometry}

If the Ramsey sequence is applied to an ac magnetic field with frequency $f\gtrsim 1/\tau$, the positive and negative contributions cancel, thus the sensitivity vanishes.
One way to avoid such an undesired cancellation is the so-called Hahn-echo sequence~\cite{Hahn:1950}, where we apply an additional $\pi$ pulse along the same axis as the first $\pi/2$ pulse at the central time $\tau/2$.
The whole sequence results in
\begin{align}
	\ket{\psi_I(\tau)} &= U_2^{\pi/2}\, U^{(2)}_{\rm free} \left(
		\frac{\tau}{2}
	\right) U_1^\pi U^{(1)}_{\rm free} \left(
		\frac{\tau}{2}
	\right) U_1^{\pi/2}\, \begin{pmatrix}
		0 \\
		1
	\end{pmatrix} \notag \\
	&= -\frac{e^{-i\Delta\phi}}{2} \left\{
		(1+ i e^{i\Delta\phi}) \ket{+} + (i + e^{i\Delta\phi}) \ket{0}
	\right\},
	\label{psi_Hahn}
\end{align}
where $\Delta\phi = \phi_2-\phi_1$, 
\begin{align}
	&\phi_1 = \int_0^{\tau/2} \gamma_e B_z(t) dt, \\
	&\phi_2 = \int_{\tau/2}^\tau \gamma_e B_z(t) dt,
\end{align}
and
\begin{align}
	&U_1^\pi =
	\begin{pmatrix}
    0 & -1 \\
    1 & 0
  \end{pmatrix},~~~~~~
  U^{(i)}_{\rm free} = \begin{pmatrix}
    e^{-i\phi_i/2} & 0 \\
    0 & e^{i\phi_i/2}
  \end{pmatrix}.
\end{align}
The signal is then estimated as
\begin{align}
	S = -\frac{1}{2}\sin(\Delta\phi).
	\label{S_Hahn}
\end{align}
One can see that for DC-like magnetic field with $f\ll 1/\tau$, $\Delta\phi\simeq 0$ and the signal vanishes. 
It also means that the effect of DC-like magnetic impurity is cancelled in this sequence, while time-dependent signals remain.
Since low-frequency noise is filtered, the relevant coherence time $T_2$ is prolonged compared to that of dc magnetometry, $T_2^{*}$, typically by one or two orders of magnitude.
Currently, the most sensitive measurement is performed with $\tau = T_2 / 2 \simeq 50\,\mathrm{\mu s}$~\cite{Wolf:2015}. Note that this transversal coherence time is the limiting factor still, as the longitudinal coherence time $T_1 \gg T_2$\cite{Jarmola:2012}.

\section{Effects of dark matter}
\label{sec:DM}

Let us apply the idea of the magnetometry in the previous section to the dark matter detection.
We consider dark matter candidates that interact with the electron spin like a magnetic field.
An example is the axion $a(\vec x,t)$, whose interaction Lagrangian and the resulting effective Hamiltonian are given by~\cite{Barbieri:2016vwg}
\begin{align}
	\mathcal L = g_{aee}\frac{\partial_\mu a}{2m_e} \bar\psi\gamma^\mu\gamma_5\psi
	~~~~~\to~~~~~ H_{\rm eff} = \frac{g_{aee}}{m_e}\vec\nabla a\cdot\vec S_e,
\end{align}
with $\vec{S}_e$ being the electron spin.
This type of interaction arises at the tree level in the DFSZ axion model~\cite{Zhitnitsky:1980tq,Dine:1981rt} or the flavorful axion models~\cite{Ema:2016ops,Calibbi:2016hwq}.
Another example is the dark photon $H_\mu(\vec x,t)$ with the kinetic mixing with the Standard Model photon~\cite{Chigusa:2020gfs},
\begin{align}
	\mathcal L = -\frac{\epsilon}{2}F_{\mu\nu} H^{\mu\nu}
	~~~~~~\to~~~~~~H_{\rm eff} = \frac{\epsilon e}{m_e}(\vec\nabla\times \vec H)\cdot \vec S_e.
\end{align}
In both cases, the effective dark matter-electron interaction Hamiltonian is expressed as
\begin{align}
	H_{\rm eff} = \gamma_e \vec B_{\rm eff}\cdot \vec S_e\, \cos(mt+\delta),
\end{align}
with
\begin{align}
	 \vec B_{\rm eff} = \sqrt{2\rho_{\rm DM}} \times
	 \begin{cases}
	 	\frac{g_{aee}}{e} \vec{v}_{\rm DM}  & {\rm for~axion},\\
		\epsilon \left(
			\vec{v}_{\rm DM} \times \hat{H}
		\right)  & {\rm for~dark~photon},
	 \end{cases}
	 \label{eq:Beff}
\end{align}
where $m$ denotes the dark matter mass, $v_{\rm DM}$ the typical dark matter velocity, $\rho_{\rm DM}$ the dark matter energy density around the Earth, $\delta$ an arbitrary phase of the dark matter oscillation, and $\hat{H} \equiv \vec{H}/|\vec{H}|$ the direction of the dark photon field.
We can estimate $\sqrt{2\rho_{\rm DM}}v_{\rm DM} \simeq 1.3\times 10^{-8}\,{\rm T}$ for $\rho_{\rm DM}=0.4\,{\rm GeV/cm^3}$ and $v_{\rm DM}=10^{-3}$~\cite{ParticleDataGroup:2022pth}.
For reference, the de Broglie wavelength of the dark matter is $\lambda = (m v_{\rm DM})^{-1} \simeq 2.0\times 10^6\,{\rm m}\,(m/10^{-10}\,{\rm eV})^{-1}$ and the coherence time is $\tau_{\rm DM}\simeq (m v_{\rm DM}^2)^{-1} \simeq 6.6\,{\rm s}\,(m/10^{-10}\,{\rm eV})^{-1}$.
As far as the de Broglie length is longer than the typical size of the diamond sample, one can regard the dark matter as a spatially uniform field.
Also within the time scale of $\tau_{\rm DM}$, one can safely approximate the dark matter field as a harmonic oscillator like $\cos(mt+\delta)$.
Below, we consider the case of $\tau < \tau_{\rm DM}$, which is satisfied for $m\lesssim 0.1\,{\rm meV}$ when $\tau\sim 1\,{\rm \mu s}$.

Note that the above expression assumes the absence of any shielding materials. 
If the experimental apparatus is shielded by a conductor with a typical size $L \ll \lambda$, $B_{\rm eff}$ for the dark photon case may be suppressed by a factor of $mL$ [$1/(mL)$] when $mL < 1$ ($mL > 1$) rather than $v_{\rm DM}$~\cite{Chaudhuri:2014dla}.
For higher frequencies with $L > \lambda$, on the other hand, the dark matter field can distinguish the detailed structure of the conductor material, and a strong dependence of the sensitivity on the material geometry is expected.
In this paper, we simply neglect such high-frequency regions for the dark photon case, leaving them as a future direction.

Time evolution of the Bloch vector is affected by the dark matter interaction with the spin triplet states. The effective Hamiltonian in the interaction picture, in the basis of $\left|+\right>$ and $\left|0\right>$, is given by
\begin{align}
	H^{\rm eff}_I &= 
	\frac{\gamma_e}{2}\cos(mt + \delta)
	\begin{pmatrix}
		B^{\rm eff}_z & \sqrt{2} B^{\rm eff}_- e^{i\omega_+t}\\
		\sqrt{2} B^{\rm eff}_+ e^{-i\omega_+t} & -B^{\rm eff}_z
	\end{pmatrix},
	\label{Heff}
\end{align}
where $B^{\rm eff}_\pm \equiv B^{\rm eff}_x \pm i B^{\rm eff}_y$. We define $t=0$ to be the injection time of the initial $\pi/2$ pulse.
Under the Ramsey sequence for dc magnetometry, the state evolves according to\footnote{
	Precisely speaking, the presence of dark matter also affects the evolution during the initial and final $\pi/2$ pulses.
	However, this dark matter effect is numerically negligible since the typical size of the magnetic fields used in the $\pi/2$ pulses is much larger than $B_{\mathrm{eff}}$.
}
\begin{align}
	 \ket{\psi_I(\tau)} = U_2^{\pi/2}\, {\rm T}\left[e^{-i\int_0^\tau H^{\rm eff}_I dt}\right]\, U_1^{\pi/2} \begin{pmatrix}
		0 \\
		1
	 \end{pmatrix},
	 \label{psi_fin_DM}
\end{align}
where T denotes the time ordering.
Note that, since typically $\gamma_e B_{\rm eff}\tau \ll 1$, one can expand $e^{-i\int_0^\tau H^{\rm eff}_I dt} \simeq {\bf I}-i\int_0^\tau H^{\rm eff}_I dt $.
Unless $m$ is very close to $\omega_+$, the off-diagonal elements of (\ref{Heff}) are rapidly oscillating.
Especially, for $m\ll 1/\tau \ll \omega_+$, only the diagonal components are important.
Then, we can calculate $S$ as
\begin{align}
	S(\delta) = \frac{1}{2}\left[\cos\theta + \sin\theta \frac{\gamma_e B_z^{\rm eff}}{m} \left(\sin(m\tau+\delta)-\sin\delta\right)\right].
\end{align}
Note that $\delta$ takes random values on time scales longer than $\tau_{\rm DM}$, although in each Ramsey sequence it is constant as far as $\tau\ll\tau_{\rm DM}$. To take account of this randomness, we define the average of an arbitrary function $f(\delta)$ as
\begin{align}
	\Braket{f} \equiv
	\frac{1}{2\pi}\int_0^{2\pi} f(\delta) d\delta.
\end{align}
Since $\Braket{S}=0$, the standard deviation $\sqrt{\Braket{S^2}}$ represents the typical size of the dark matter signal.
For $\theta=\pi/2$, it is calculated as
\begin{align}
	\sqrt{\Braket{S^2}} = \frac{\gamma_e B_z^{\rm eff}}{2m}\sqrt{1-\cos(m\tau)}.
 \label{S_fin_DM}
\end{align}
In the small mass limit $m\tau\ll 1$, this gives $\sqrt{\Braket{S^2}}\simeq \phi/\left( 2\sqrt 2 \right)$, which is analogous to the case of a dc magnetic field in Eq.~\eqref{S_fin_DC} as expected.

The time variation of the dark matter induced magnetic field becomes important when $m \tau \gtrsim 2\pi$ or $m \gtrsim 10^{-10}\,\mathrm{eV}$.
In this case, it is more efficient to resort to the Hahn-echo sequence for ac magnetometry.
From Eq.~(\ref{S_Hahn}), by taking $B_z(t) = B_z^{\rm eff}\cos(mt+\delta)$, we obtain
\begin{align}
	S(\delta) \simeq \frac{2\gamma_e B_z^{\rm eff}}{m} \sin^2 \left(
		\frac{m\tau}{4}
	\right) \sin\left(
		\frac{m\tau}{2}+\delta
	\right) ,
\end{align}
which yields $\Braket{S}=0$, and the standard deviation
\begin{align}
	\sqrt{\Braket{S^2}} =
	\frac{\sqrt{2}\gamma_e B_z^{\rm eff}}{m} \sin^2 \left(
		\frac{m\tau}{4}
	\right).
	\label{eq:S_DM_ac}
\end{align}

\section{Sensitivity of dc magnetometry}
\label{sec:dc}

In NV center magnetometry, the phase of its electron spin depends on the magnetic field strength. Some final spin manipulations allow to convert this change in phase to a signal $S$ defined in Eq.~\eqref{S_fin_DC}.
In this section, using results from Sec.~\ref{sec:magnetometry} and \ref{sec:DM}, we show the sensitivity of dc magnetometry with NV centers on light bosonic dark matter.

In the most optimistic setup, the unique noise source is the intrinsic quantum fluctuation of the spin, which is called the spin projection noise.
It gives an inevitable contribution to $S$ in an ensemble magnetometer represented as
\begin{align}
	\Delta S_{\mathrm{sp}} = \frac{1}{2}
	\frac{1}{\sqrt{N \left( t_{\mathrm{min}} / \tau \right)}},
	\label{eq:sp-noise}
\end{align}
where $N$ is the number of NV centers, $t_{\mathrm{min}} \equiv \min \left( t_{\mathrm{obs}}, \tau_{\mathrm{DM}} \right)$ with $t_{\mathrm{obs}}$ and $\tau_{\mathrm{DM}}$ being the total observation time and the coherence time of dark matter, respectively, and $\tau$ is the free precession time.
This is for an ensemble of independent NV centers, thus the resource of quantum entanglement is not utilised yet.
When the spin projection noise is the dominant noise source, the Ramsey sequence is sensitive to a magnetic field as weak as\footnote{
In this expression, we neglect the possible sensitivity loss from the imperfect readout, overhead time, and shot noise (see, e.g., \cite{Barry:2019sdg}) for simplicity of the expression.
For a typical size of overhead time $t_o \sim O(T_2^{*})$, this could affect the sensitivity by $O(1)$ through the replacement $\tau \to \tau + t_o$ in Eq.~\eqref{eq:sp-noise}.
In the sensitivity plots, we plot the spin-projection noise limited sensitivities, and as comparison for the magenta dashed lines also the sensitivities of the current state-of-the-art experimental sensor, which thus includes all sensitivity losses.
}
\begin{align}
	\Delta B_{\rm sp} &\simeq \frac{e^{\tau/T_2^{*}}}{\gamma_e \sqrt{N \tau t_{\rm min}}}
	\simeq 13\,{\rm fT}\left(\frac{10^{12}}{N}\right)^{1/2}\left(\frac{0.5\,{\rm \mu s}}{\tau}\right)^{1/2}
	\left(\frac{1\,{\rm s}}{t_{\rm min}}\right)^{1/2},
	\label{eq:proj_sensitivity}
\end{align}
The exponential factor represents the sensitivity loss according to the spin dephasing with a relaxation time $T_2^{*}$.
This factor makes $\tau\sim T_2^*/2$ to be the optimal choice to maximize the sensitivity~\cite{Herbschleb:2019}, which is assumed on the right-hand side.

\begin{figure}
	\centering
		 \includegraphics[width=1.0\hsize]{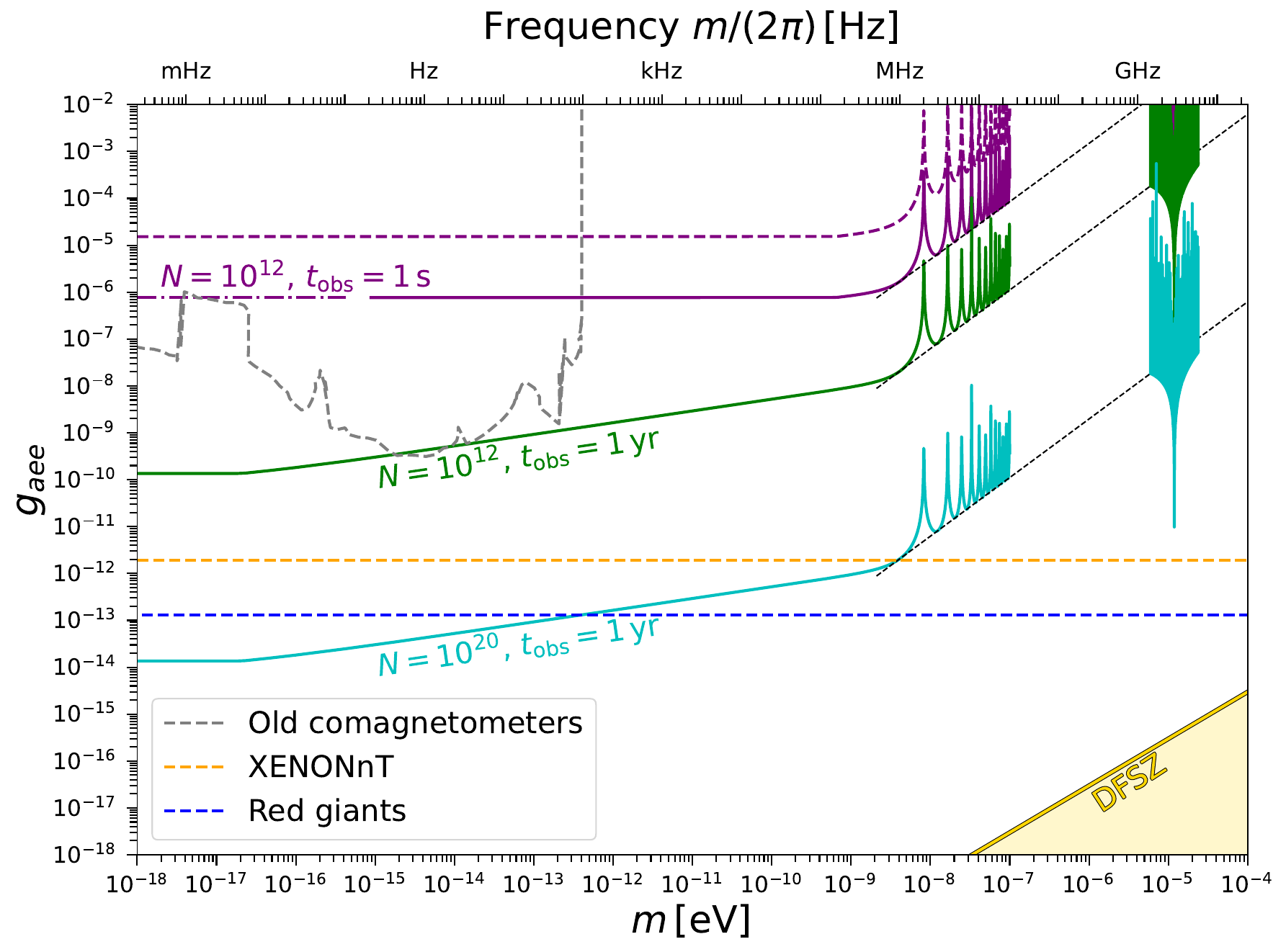}
			\includegraphics[width=1.0\hsize]{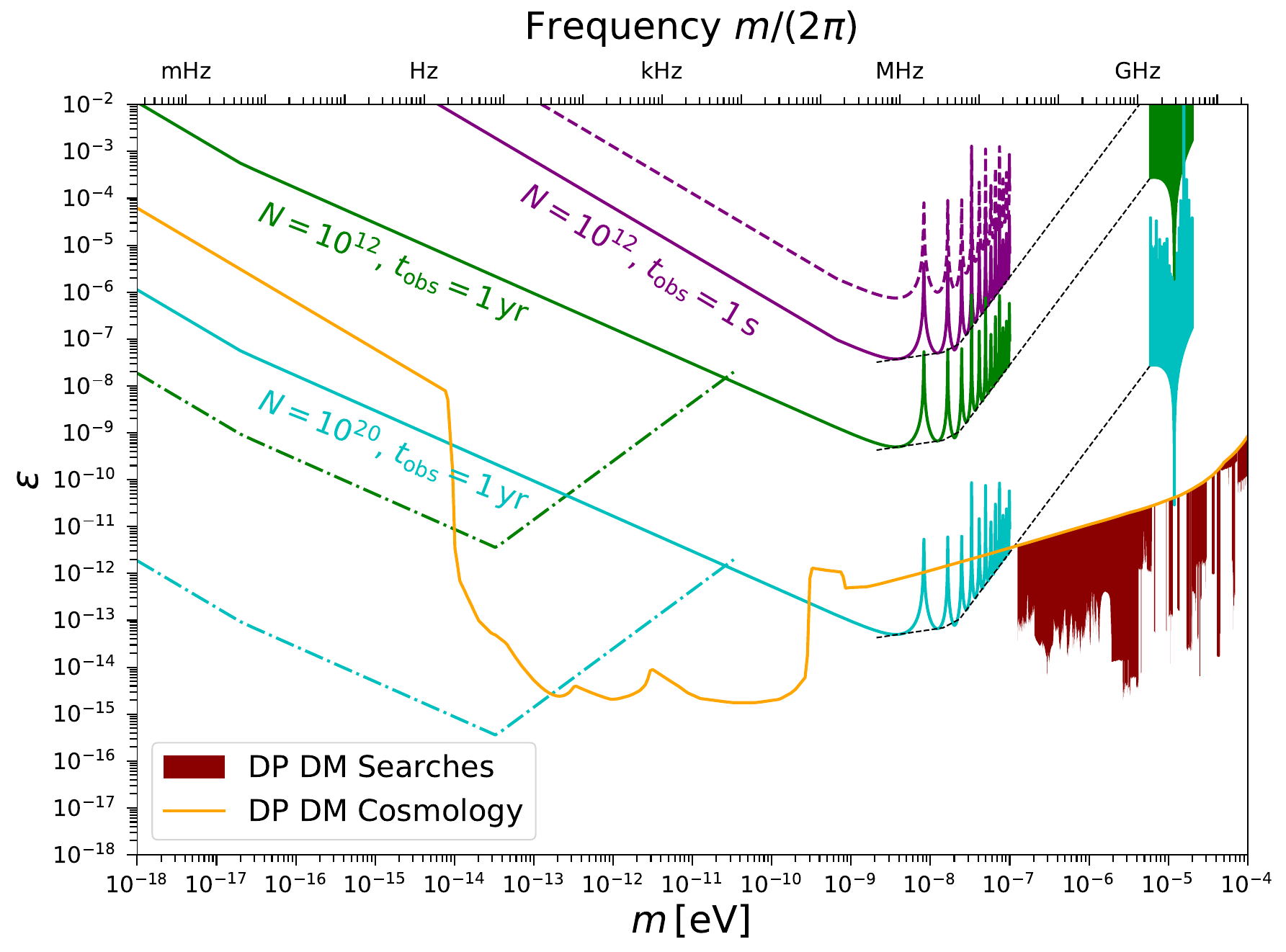}
	 \caption{The reach of diamond NV-center magnetometry on dark matter models. (Top) The case of axion dark matter coupling with electron. (Bottom) The case of dark photon dark matter with kinetic mixing with the ordinary photon.
	 The colored solid (dash-dotted) lines correspond to the case with (without) the magnetic shielding.
	 Note that the dc magnetometry (lighter region) and the resonance search (GHz/heavier region) require different sequences and cannot be performed simultaneously; see the text for details.
	 The black dashed lines show tangent lines of the sensitivity curves to guide the eye.
	 }
	 \label{fig:dm}
 \end{figure}

If $t_{\mathrm{obs}} > \tau_{\mathrm{DM}}$, repeated measurements allow us to extract the possible dark matter signal as broadening of the width of the observed signal distribution.
$\Braket{S^2}$ calculated in Eq.~\eqref{S_fin_DM} can be used as the dark matter contribution to the width, which is the ensemble average of the dark matter signal squared.
Accordingly, the sensitivity on dark matter models is estimated by solving
\begin{align}
    \sqrt{\Braket{S^2} e^{-2\tau/T_2^*} + (\Delta S_{\mathrm{sp}})^2} - \Delta S_{\mathrm{sp}} \gtrsim
    \frac{\Delta S_{\mathrm{sp}}}{\sqrt{2 t_{\mathrm{max}} / \tau_{\mathrm{DM}}}},
\end{align}
with $t_{\mathrm{max}} \equiv \max \left( t_{\mathrm{obs}}, \tau_{\mathrm{DM}} \right)$.
In the above inequality, the left-hand side represents the dark matter effect on the width of the signal distribution, while the right-hand side the estimation uncertainty associated with the unbiased estimation of standard deviation when we repeat the measurement $t_{\mathrm{max}} / \tau_{\mathrm{DM}}$ times.
Combined with Eq.~\eqref{eq:sp-noise}, this gives us the sensitivity
\begin{align}
    \sqrt{\Braket{S^2}} e^{-\tau/T_2^{*}} \gtrsim \frac{1}{2^{3/4}}
    \frac{1}{\sqrt{N \left( t_{\mathrm{min}} / \tau \right)}}
    \frac{1}{\left( t_{\mathrm{max}} / \tau_{\mathrm{DM}} \right)^{1/4}}.
\end{align}
To take into account the randomness of the direction and amplitude of $\vec{B}_{\mathrm{eff}}$ given by Eq.~\eqref{eq:Beff}, we replace $\left( B_z^{\mathrm{eff}} \right)^2$ by $B_{\mathrm{eff}}^2/3$ and use a typical dark matter velocity $v_{\mathrm{DM}}=10^{-3}$.

The resulting sensitivities on the axion-electron coupling $g_{aee}$ and the kinetic mixing parameter of the dark photon $\epsilon$ are shown in Fig.~\ref{fig:dm}. We draw three solid lines for three experimental setups: $(N,t_{\rm obs}) = (10^{12}, 1\,{\rm s}),  (10^{12}, 1\,{\rm year}),  (10^{20}, 1\,{\rm year})$ with $\tau=0.5\,\mathrm{\mu s}$.
The number of NVs $N=10^{12}$ for the first two sets is attainable given previous experiments~\cite{Barry:2023hon}.
Obtaining $N = 10^{20}$ requires a volume of approximately $10^3\,{\rm cm^3}$ for a NV concentration of $1.6 \times 10^{17}{\rm cm^{-3}}$ in diamonds with high sensitivity~\cite{Wolf:2015,Barry:2023hon}. The large volume can be obtained using a synthesis technique for a large diamond~\cite{Schreck:2017}, in which case a laser power of $5\times10^{2}$~kW would be required in the current technique~\cite{clevenson2015broadband}, and sufficient microwave homogeneity might be challenging to attain.
Alternatively, the same sensitivity can be obtained by combining smaller individual sensors, which decreases the laser power and the microwave inhomogeneity in the individual sensors.
For example, with similar growth as for the high-sensitivity diamond with a thickness of $70\,\mathrm{\mu m}$~\cite{Barry:2023hon}, using large diamond wafers~\cite{Schreck:2017}, in the order of $10^3$ wafers are required to reach the required volume, a number that can be reduced with future improvements such as increasing yield and thickness.
The sensitivities of the dc magnetometry drastically oscillate for heavier masses, so we do not plot them for $m> 10^{-7}\,\mathrm{eV}$ and show their tangent lines instead.
Together shown are constraints from red giant stars~\cite{Capozzi:2020cbu}, solar axion search in XENONnT~\cite{XENON:2022ltv}, cosmological bounds on the dark photon~\cite{McDermott:2019lch,Witte:2020rvb},
and dark photon dark matter search constraints~\cite{Caputo:2021eaa, Brun:2019kak, Godfrey:2021tvs, Nguyen:2019xuh, Dixit:2020ymh, Cervantes:2022yzp, DOSUE-RR:2022ise, An:2022hhb, An:2023wij, Ramanathan:2022egk}.
The constraint from the old comagnetometers on $g_{aee}$~\cite{Bloch:2019lcy} is also shown for the purpose of comparison.
See Ref.~\cite{OHare:2020} for a summary of existing constraints on the axion and the dark photon.
The yellow band shows the DFSZ axion model \cite{Zhitnitsky:1980tq,Dine:1981rt} under the constraint of $0.28 < \tan\beta < 140$~\cite{Chen:2013kt}.
Although the currently reported $\Delta S$'s are worse than $\Delta S_{\mathrm{sp}}$ by roughly a factor of $20$, it can approach $\Delta S_{\mathrm{sp}}$ in principle by optimizing the experimental setup, for example, improving read-out fidelity such as photon collection efficiency and suppression of noises from equipment~\cite{Barry:2019sdg}.
For comparison, the shot-noise-limited sensitivities for the setup $(N,t_{\rm obs}) = (10^{12}, 1\,{\rm s})$ are also shown by the magenta dashed lines in Fig.~\ref{fig:dm}.

If $\tau_{\mathrm{DM}} \ll t_{\mathrm{obs}}$, effective magnetic fields with different amplitudes and phases contribute during the measurement, and the observation result of each Ramsey sequence distributes with an average $\Braket{S}=0$ and a standard deviation $\sqrt{\Braket{S^2}}$.
On the other hand, if $\tau_{\mathrm{DM}} \gg t_{\mathrm{obs}}$, both the amplitude and phase are fixed during the whole measurement duration.
Even in this case, the directional dependence on $\vec{B}_{\mathrm{eff}}$ of the sensitivity can be averaged by, e.g., using different sets of NV centers with different axis directions.\footnote{
	The orientation of NV centers in the diamond sample can be aligned~\cite{Fukui:2014,Miyazaki:2014,Michl:2014,Lesik:2014}.
} 
Also, if $1/m \ll t_{\mathrm{obs}} \ll \tau_{\mathrm{DM}}$, the overall cosine factor oscillates over successive Ramsey sequences, resulting in the same distribution of $S$.
However, when $1/m \gg t_{\mathrm{obs}}$, which is shown by dash-dotted lines, a randomly sampled phase sets the maximum possible sensitivity of the measurement to dark matter mass, which can even make the measurement insensitive at all.
In this case, the plotted sensitivity should be interpreted as an upper bound at the $68\%$ confidence level.
Finally, when $1/m \gg \tau$, the dc magnetometry approach loses sensitivity due to the signal oscillation.
Note, however, that the blind spots at which the plot shows zero sensitivity are merely an artifact of the specific choice of $\tau=0.5\,\mathrm{\mu s}$, and can be covered by repeating the measurement with slightly different choices of $\tau$.

Compared to the axion signal that could be totally unaffected by the magnetic shielding~\cite{JacksonKimball:2016wzv}, the dark photon signal is inevitably affected.
To show the effect of the magnetic shielding in the bottom panel of Fig.~\ref{fig:dm}, we plot both results with and without magnetic shielding with the colored solid and dash-dotted lines, respectively.
For the former case, we assume the typical size of the magnetic shielding made of a conducting material with a typical size of $L\sim 10\,\mathrm{m}$.
This setup is the most suitable to explore the frequency with $mL\sim 1$, i.e. $m\sim 2\times 10^{-8}\,\mathrm{eV}$ or $m/(2\pi) \sim 5\,\mathrm{MHz}$.
Note that the resulting magnetic field around this frequency is larger than the naive expectation \eqref{eq:Beff} by a factor of $v_{\rm DM}^{-1}$.
This is due to the fact that the magnetic field detected here is sourced from the oscillation of electrons inside the shielding material, which is caused by the electric field component of the dark photon oscillation unsuppressed by $v_{\rm DM}$.

With larger magnetic shielding, the sensitivity peak for the kinetic mixing parameter $\epsilon$ shifts towards a smaller frequency.
In the end, without any artificially equipped magnetic shielding, the sensitivities represented by the dotted lines are expected, which are shown only for green and cyan setups for readability of the plot.
The absence of a shield implies that we effectively have $L\sim L_E = 6\times 10^{3}\,\mathrm{km}$, which is the Earth's radius that is proven to characterize the shielding effect from the conducting material inside the Earth and the ionosphere~\cite{Dubovsky:2015cca, Fedderke:2021aqo}.
The sensitivity lines stop at the frequency that satisfies $L_E = \lambda$, above which a strong dependence on the geometry of the conducting material is expected.
To achieve the corresponding sensitivities, however, we need some techniques such as the one introduced in \cite{Herbschleb:2022jty} to make the measurement without the magnetic shielding possible by distinguishing the oscillating signal from any other dc-like noises. 
The idea of this technique is that while any dc-like noise averages out, the long coherence time at low masses allows multiple periods with many data points per period to be accumulated to accentuate the small signal, similar to their $1$~Hz result.
This technique is applicable for a higher frequency than $1/t_{\mathrm{obs}}$ so that the oscillation is observable, which is satisfied for both the green and cyan lines.
Note that the possible existence of low-frequency noises limits the sensitivity of the search without shielding at the corresponding dark photon mass, so the understanding of the environmental magnetic fields is an important future task even with this technique.

\section{Resonance search}
\label{sec:res}

So far, we considered the off-resonant regime. If the dark matter mass $m$ is very close to the energy gap between two spin states, $\omega_+$, resonance occurs. Interestingly, $\omega_+\sim 10\,{\rm \mu eV}$ is a motivated mass range for the axion, so we comment on this regime. The resonance enhancement happens for the case of $|\Delta \omega \tau| \ll 1$ where $\Delta\omega \equiv \omega_+-m$, which implies $|\Delta\omega/\omega| \ll (\omega_+\tau)^{-1} \simeq 6\times 10^{-5} (1\,{\rm \mu s}/\tau)$. Thus, it is necessarily a narrow band search in contrast to the broad band off-resonance search.

Let us evaluate the interaction Hamiltonian (\ref{Heff}) for $\omega_+ \simeq m$. 
It is rewritten as
\begin{align}
	H^{\rm eff}_I = 
	\frac{\gamma_e}{4}
	\begin{pmatrix}
		2B^{\rm eff}_z\cos(mt + \delta) & X \\
		X^* & -2B^{\rm eff}_z \cos(mt + \delta)
	\end{pmatrix},
\end{align}
where
\begin{align}
	X = \sqrt{2} B^{\rm eff}_- \left[ e^{i(\omega_+-m)t-i\delta} + e^{i(\omega_+ +m)t+i\delta} \right].
\end{align}
After integrating the Hamiltonian over time, only terms dependent on $\Delta\omega$ are important. This results in
\begin{align}
	e^{-i\int_0^\tau H^{\rm eff}_I dt} \simeq \begin{pmatrix}
		1 & Y \\
		-Y^* & 1
	\end{pmatrix},
\end{align}
where
\begin{align}
	Y = -i\frac{\sqrt{2}}{4} \gamma_e B_- \tau e^{-i\delta}.
\end{align}
In this case, we need a new sequence that is similar to the Ramsey sequence but without the second $\pi/2$ pulse to extract the dark matter effect at the linear order. The final state is
\begin{align}
	 \ket{\psi_I(\tau)} = {\rm T}\left[e^{-i\int_0^\tau H^{\rm eff}_I dt}\right]\, U_1^{\pi/2} \begin{pmatrix}
		0 \\
		1
	 \end{pmatrix}
	 = \frac{1}{\sqrt 2}  \begin{pmatrix}
		Y-1 \\
		Y^*+1
	 \end{pmatrix}.
\end{align}
Then we obtain $\Braket{S}=0$ after average over $\delta$, and
\begin{align}
	\sqrt{\Braket{S^2}} 
	= \frac{\gamma_e}{2\sqrt{2} \left|\Delta\omega\right|}\sqrt{1-\cos(\Delta\omega\tau)}\sqrt{\left(B_x^{\rm eff}\right)^2+\left(B_y^{\rm eff}\right)^2}.
	\label{S_fin_res}
\end{align}
Compared with the off-resonance estimation Eq.~\eqref{S_fin_DM}, this is enhanced by a factor of $m\tau \simeq 2\times 10^4 (\tau/1\,{\rm \mu s})$ at the resonant point $|\Delta \omega \tau| \ll 1$. One can scan the dark matter mass by changing the bias magnetic field $B_0$ so that $\omega_+$ changes. Although we focused on the $\left|+\right>$ state, all the discussion is almost parallel for the case of the $\left|-\right>$ state and hence the resonance also happens for the dark matter mass $m\simeq \omega_-$.
We plot in Fig.~\ref{fig:dm} sensitivities for the resonance case.

\section{Sensitivity of ac magnetometry}
\label{sec:ac}

\begin{figure}[t]
  \centering
  \includegraphics[width=\hsize]{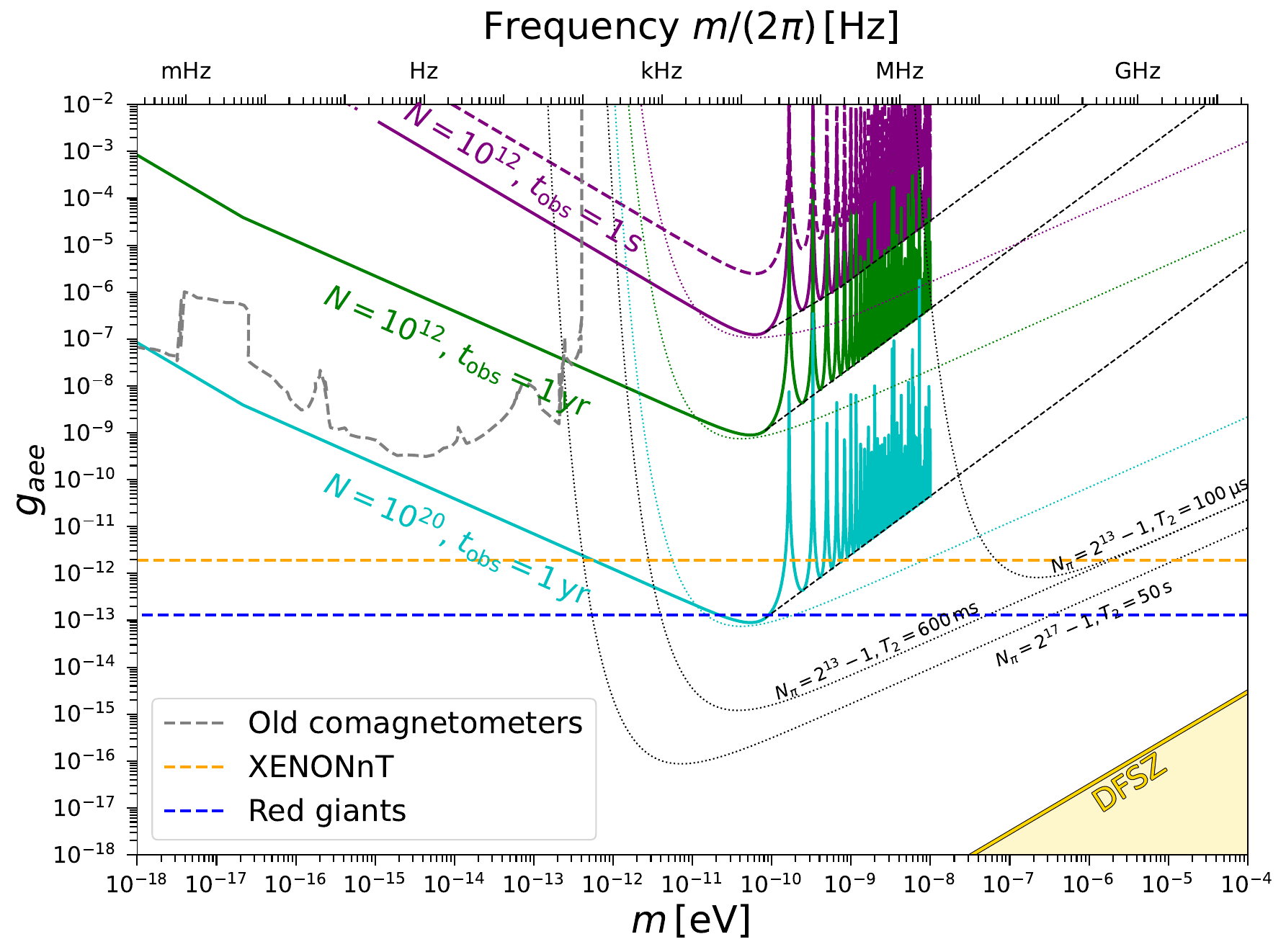}
	\includegraphics[width=1.0\hsize]{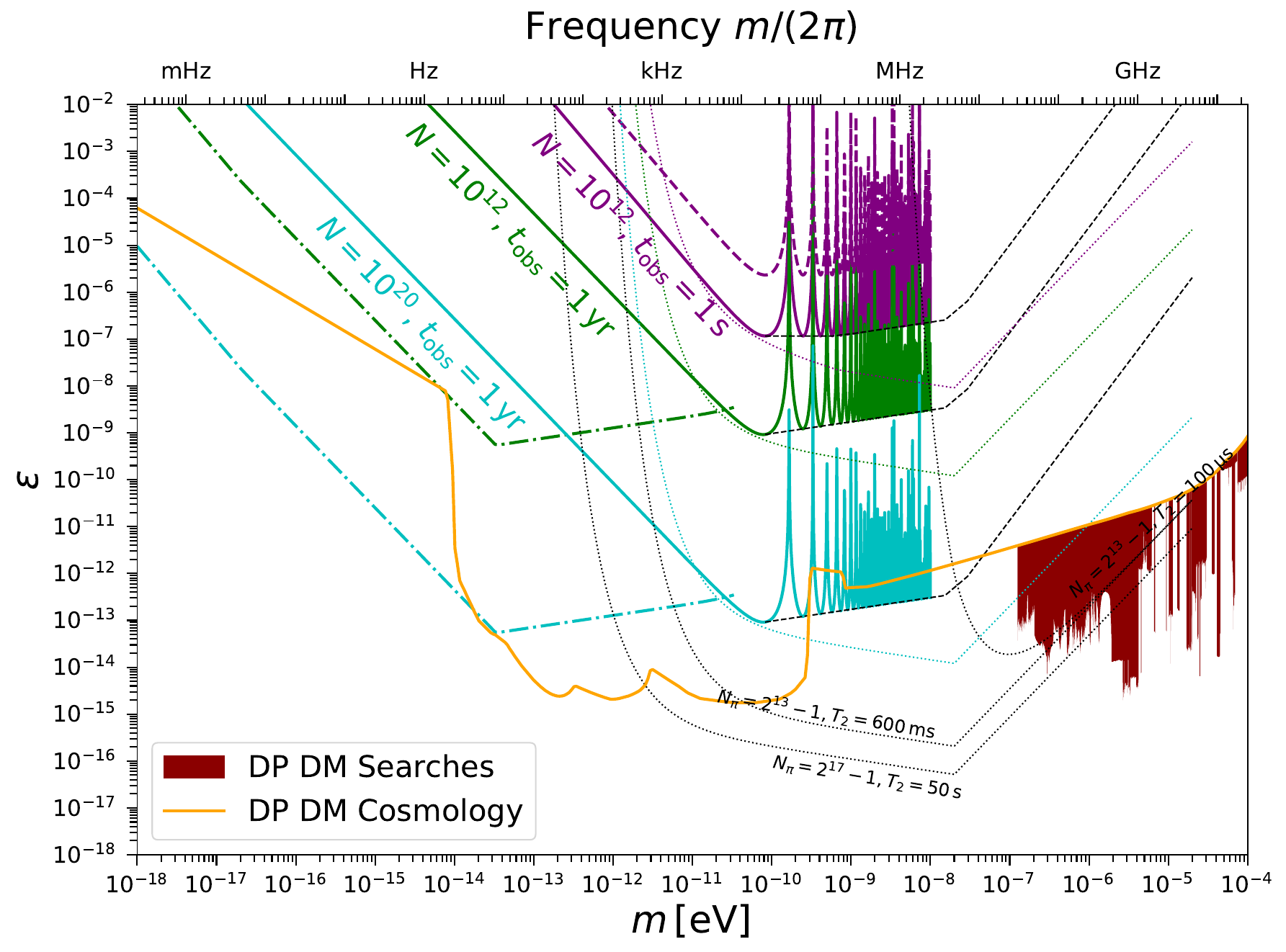}
  \caption{
		The reach of diamond NV-center ac magnetometry on dark matter models.
		The setups and the color conventions are the same as those in Fig.~\ref{fig:dm}.
		The colored and black dotted lines denote the possible reach when we use different choices of $\tau$ and different pulse sequences; see the text for details.
	}
  \label{fig:reach_ac}
\end{figure}

For ac magnetometry, the dark matter contribution to the width of the signal distribution is calculated in Eq.~\eqref{eq:S_DM_ac}.
Similarly to dc magnetometry, the sensitivity is obtained by solving $\sqrt{\Braket{S^2}} e^{-\tau/T_2} / \Delta S_{\mathrm{sp}} = 1$, where $T_2$ is the coherence time.
The resulting constraints are shown in Fig.~\ref{fig:reach_ac}.
It is worth noting that the sensitivity to dark matter mass around the target angular frequency $2\pi/\tau$ is better than that of the dc magnetometry approach thanks to the longer coherence time $T_2 \gg T_2^{*}$.

Moreover, we can design experiments that focus on higher frequencies. The sensitivities by optimizing $\tau$~\cite{Herbschleb:2019} for each mass are plotted with colored dotted lines in Fig.~\ref{fig:reach_ac}.
They are envelopes of sensitivity curves with different choices of $\tau$ and characterize the wide dynamic range of our approach.
Besides, the higher mass region becomes more accessible by increasing the number of $\pi$ pulses $N_\pi$ with so-called dynamical decoupling sequences~\cite{Meiboom:1958}. Finally, as the coherence time for the decoupling sequence is limited by $T_1$, which increases by orders of magnitude at low temperatures~\cite{Jarmola:2012}, the bounds towards the lower masses could be improved.
The three black dotted lines in Fig.~\ref{fig:reach_ac} show the envelopes for $N=10^{20}$ and $t_{\mathrm{obs}}=1\,\mathrm{yr}$ to demonstrate how the sensitivity changes with different setups for the following parameters. The line with $(N_\pi; T_2)=(2^{13}-1;100\,\mathrm{\mu s})$ illustrates that the more pulses, the higher the mass at the optimum. 
The line with $(2^{13}-1;600\,\mathrm{m s})$ corresponds to the currently known best $T_2\approx 0.6$~s at $77$~K~\cite{BarGill:2013}. 
The line with $(2^{17}-1;50\,\mathrm{s})$ is based on the current longest $T_1$~\cite{Jarmola:2012}. 
Sample advancements could improve this further.
Note that these sensitivity lines are envelopes of each narrow band search for large $N_\pi$. In one experimental setup, the sensitivity has a narrow peak. In this sense, these lines should be regarded as an optimal sensitivity at each mass.

\section{Conclusions and discussion}
\label{sec:con}

We proposed a novel method to search for light dark matter, such as the axion or the dark photon, by utilizing the NV-center magnetometry method.
As shown in Figs.~\ref{fig:dm} and \ref{fig:reach_ac}, the projected reach can go beyond the current experimental and observational limits.
In order to reach the prediction on $g_{aee}$ from the axion for solving the strong CP problem~\cite{ParticleDataGroup:2022pth}, we may still require a few orders of magnitude improvement.
Note again that dotted lines in Fig.~\ref{fig:reach_ac} are envelopes of each narrow band search, representing an optimized sensitivity in each target mass. 
One possibility for the improvement is to have even longer $T_2$, which may be achieved by cooling the system~\cite{Jarmola:2012} and using a large number of pulses~\cite{Meiboom:1958}. The sensitivity may also be greatly enhanced further by using entangled quantum states~\cite{Degen:2017}.

For the axion dark matter case, although we focused on the axion-electron coupling $g_{aee}$ in the main text, our setup also has a good sensitivity to the axion-photon coupling $g_{a\gamma\gamma}$, since a bias magnetic field $\vec B_0$ is applied to the diamond sample. One can naively convert the sensitivity on $g_{aee}$ to $g_{a\gamma\gamma}$ through $g_{a\gamma\gamma} \simeq mg_{aee}/(eB_0) \sim g_{aee}\times17\,{\rm GeV^{-1}}(1\,{\rm G}/B_0)(m/10^{-10}\,{\rm eV})$.\footnote{
    If the orientation of the NV center ensembles and $\vec B_0$ are perfectly aligned to the same axis, the sensitivity on $g_{a\gamma\gamma}$ would be reduced since the axion-induced magnetic field is proportional to $\vec v_{\rm DM}\times \vec B_0$.
}

In this paper, we focus on the theoretical idea to exploit NV centers for dark matter search, since the best experimental sensitivities keep improving. Nonetheless, we would like to compare our theoretical results with the state-of-the-art at time of writing. Compared to the best reported sensitivities \cite{Barry:2023hon}, the investigated quantum-projection-noise sensitivity is about $20$ times better, thus moving the lines in the figures roughly an order of magnitude up, as indicated for the setup $(N,t_{\rm obs}) = (10^{12}, 1\,{\rm s})$ with the magenta dashed lines in Figs.~\ref{fig:dm}~and~\ref{fig:reach_ac}. Note that these experimental sensitivities include the consequences of waiting times, initialization/readout laser pulses and their noise, microwave pulses and their noise; thus, all experimental factors are included for the purple dashed line in all plots. In recent years, the sensitivity has improved by over an order of magnitude \cite{Wolf:2015,Barry:2023hon}, and we expect it to increase further still in the years to come.

Optically-pumped magnetometry (OPM) \cite{Budker:2007} may similarly realize the detection of dark matter in principle. However, the measurable region of the NV center without a magnetic shield is wider than that of OPM due to its wider dynamic range \cite{Budker:2007, Herbschleb:2021}. It is an advantage of the NV center for the detection of the dark photon, because a conductor material suppresses the sensitivity to detect it \cite{Chaudhuri:2014dla}.

Finally, we mention comparisons with the light dark matter detection using magnons. Magnons are collective excitations of the electron spin in a magnetic material and it has a gap energy corresponding to the Larmor frequency under an external magnetic field. Refs.~\cite{Barbieri:2016vwg,Chigusa:2020gfs} considered a resonant amplification of the magnon when the axion or dark photon mass is equal to the magnon gap energy, and hence it is a narrow band search and sensitive to the mass range around $10^{-4}$\,eV. On the other hand, for our methods in this letter, very broad mass range can be searched for by using the magnetometry technique.


\begin{acknowledgments}
We would like to thank Hideo Iizuka for useful discussion. KN would like to thank Wen Yin for comments.
This work was supported by the Director, Office of Science, Office of High Energy Physics of the U.S. Department of Energy under the Contract No. DE-AC02-05CH1123 [SC].
This work was supported by JSPS KAKENHI Grant (Nos. 18K03609 [KN] and 17H06359 [KN]).
This work was partially supported by MEXT Q-LEAP (No. JPMXS0118067395 [NM, EDH]).
This work was supported by World Premier International Research Center Initiative (WPI), MEXT, Japan.
\end{acknowledgments}

\bibliographystyle{apsrev4-1}
\bibliography{bib}

\end{document}